\documentclass[aps,twocolumn,superscriptaddress,floatfix]{revtex4}

\usepackage{graphicx}
\usepackage{amsmath}

\begin{document}

\title{Network effects in service usage}

\author{G\'abor Szab\'o}
\affiliation{Center for Complex Network Research and Department of Physics,
University of Notre Dame, IN 46556, USA}
\affiliation{Center for Cancer Systems Biology and Department of Cancer
Biology, Dana--Farber Cancer Institute and Department of Genetics,
\hspace{2cm} Harvard Medical School, Boston, MA 02115, USA}

\author{Albert-L\'aszl\'o Barab\'asi}
\affiliation{Center for Complex Network Research and Department of Physics,
University of Notre Dame, IN 46556, USA}
\affiliation{Center for Cancer Systems Biology and Department of Cancer
Biology, Dana--Farber Cancer Institute and Department of Genetics,
\hspace{2cm} Harvard Medical School, Boston, MA 02115, USA}

\begin{abstract}
While there is ample evidence that social and communication networks
play a key role during the spread of new ideas, products, or services,
network effects are expected to have diminished influence in the
stationary state, when all users are aware of the innovation, and its
usage pattern is determined mainly by its utility to the user. Here we
study four mobile phone-based services available to over six million
subscribers, allowing us to simultaneously monitor the communication
network between individuals and the time-resolved service usage
patterns. We find that usage highly correlates with the structure of
the communication network, and demonstrate the coexistence on the same
social network of two distinct usage classes, network effects being
responsible for the quantifiable differences between them. To test the
predictive power of our theory, we demonstrate that traditional
marketing techniques are ineffective in permanently boosting service
adoption, and propose a hub-based incentive mechanism that has the
potential to enhance usage for one of the two service classes.
\end{abstract}

\maketitle

Thanks to electronic communications that offer virtually instantaneous
information access, awareness of some products and services, such as
iPod, MySpace, RAZR, and ringtones, have swept the population within a
few months, resulting in millions of active users. Innovation
diffusion studies offer a rather complete conceptual picture of this
adoption process, capturing how an innovation spreads from early
adopters to the laggards along the links of the social and
communication network
\cite{rogers,bass,valente,mahajan,helbing,guardiola}. Yet, all
innovations eventually reach a saturated state, in which there is
complete awareness of its benefits, and practically all those who
could find the innovation useful have adopted it. The social network,
that plays a key role during the spreading process \cite{brancheau},
is not expected to impact this saturated state, as only the utility of
the product or service to the individual user determines its
day-to-day usage. This leaves open a question of major conceptual and
practical importance: could the underlying social and communication
network play a role in the usage of innovations? Here we show that
this is indeed the case: the ties between individuals significantly
impact the usage patterns of many services.

Today our ability to study innovation diffusion and usage patterns is
enhanced by the fact that virtually all electronic communications and
purchasing events leave digital signatures
\cite{eckmann,bornholtz,barabasi}.  Yet, advances are still limited by
the fact that we need to \emph{simultaneously} track both social
interactions and service usage patterns, which requires the merging of
various proprietary databases, a procedure limited by significant
technical and legal roadblocks, as well as legitimate confidentiality
concerns. Given that mobile carriers must track for billing purposes
each call users place or receive, effectively recording the underlying
social network, and regularly introduce innovative phone-based
services, they offer unprecedented opportunities to explore the
detailed dynamics of service usage \cite{vesa}. Here we capitalize on
this opportunity, studying the role of the social network in the usage
patterns of four mobile phone-based services, available to over a
quarter of mobile phone users in a country. We first show that service
usage is characterized by detectable dyadic effects, indicating that
if a user uses a specific service in a regular fashion, her
phone-based contacts will also use it regularly. Yet, despite this
dyadic effect, we find that the four services cluster into two
distinct classes, documenting the coexistence on the same social
network of two rather different usage patterns. We develop an
analytical theory, rooted in diffusion models used regularly to model
spreading effects in physical
sciences~\cite{havlin,matteo,bettencourt} and the spread of computer
and biological
viruses~\cite{berthelemy,vespignani-virus,vespignani-internet} to
predict the role of the social network in service usage, and
demonstrate that it accurately describes the observed usage
dynamics. To test the predictive power of the proposed quantitative
framework, we explore the impact of advertising on service usage,
showing that for the studied services advertising has no long-term
impact, a prediction that is directly supported by real data. Finally,
we propose that hub-based incentives could be used to permanently
boost service usage, but they are effective for one service class
only, illustrating the practical importance of identifying the right
class to which a service belongs to.

The dataset studied by us records all calls initiated by over 6
million customers of a mobile phone company, allowing us to represent
each phone number, replaced by a hash code to guarantee anonymity, as
a node in the communication network. The undirected links between
users represent voice calls or short messages, the two most widespread
mobile phone-based communication services, resulting in approximately
10 million links each month. The call data spans a period of over a
year, and is aggregated into intervals of two weeks, each record
identifying the call's origin, destination, and the type of service
used by the user. We focus on four services: Email, whose users can
send and receive emails from their cell phones; WAP, that gives the
user the opportunity to browse Web sites; Chat, that allows users to
exchange short messages anonymously; and MyFriends, which allows a
group of registered users to contact all other members of the group at
once.

To characterize the service usage we first measured the number of
active users for each service for over a year-long period. We find
that the number of users who have used the service at least once from
the beginning of our observational period increases monotonically
(Fig.~1\emph{a}--\emph{d}), suggesting the presence of a classical
adoption process, during which each month new users adopt each
service. Yet, the number of individuals who use the service in a
two-week interval is largely independent of time
(Fig.~1\emph{a}--\emph{d}).  Therefore, these services have swept the
current user base and reached a saturated or endemic state before the
beginning of our observational period. The increasing cumulative user
number is not an evidence of a continued adoption process, but
indicates that the saturated state hides a highly dynamic equilibrium,
where the adopters use each service only sporadically. Indeed, the
percentage of users that use the same service in two consecutive
months is relatively small (37\% for Chat, 30\% for Email, 47\% for
WAP, and 33\% for MyFriends). Thus, in contrast with the traditional
subjects of innovation diffusion studies, where once an individual
accepts an innovation she is considered an adopter for the rest of the
study, mobile phone-based services display a highly volatile usage
pattern, with users entering and exiting the user base in large
numbers in each two week period \cite{keaveney}. This may be rooted in
the fact that these services are billed only for the air-time occurred
during their usage, thus turning the specific service on or off does
not require additional financial commitments. Thus the only lock-in
incentives are utility or network-based: a service may be
indispensable for an individual or there may be network externalities,
resulting from peer pressure by social contacts towards its use.

The fundamental hypothesis of all adoption studies is that information
externalities play a crucial role in the spreading process: a given
innovation spreads along the social links through advice, influence,
or persuasion \cite{rogers,bass,valente}. Yet, in the saturated state,
when the user base is largely aware of the benefits of an innovation,
social ties cease to exert a role as information source, thus usage
should be mainly utility-based. To establish if personal contacts
affect the usage pattern we determined the probability $\rho_i$ that
two users that communicate with each other use the same service
$S_i$. By itself $\rho_i$ is not informative, as with probability
$\rho_i^{\mathrm{rand}} = p_i^2$ any two individuals may also use the
same service, where $p_i$ represents the fraction of mobile phone
users that adopted $S_i$. Thus a better measure of the role of
personal recommendation is $\rho_i / \rho_i^{\mathrm{rand}}$. We find
that for Chat and MyFriends $\rho_i / \rho^{\mathrm{rand}} \gg 10$,
while for Email and WAP $\rho_i / \rho^{\mathrm{rand}} \approx 2$ (see
Fig.~7 in the \emph{Supporting Text}), evidence of a dyadic
effect. Therefore, users that know each other are more likely to use
the same service than expected based on a random adoption model. Thus
the impact of the communication ties is not limited to the adoption
phase, but it also affects the day-to-day usage of services.

\begin{figure*}
\begin{center}
\includegraphics[width=0.8\linewidth]{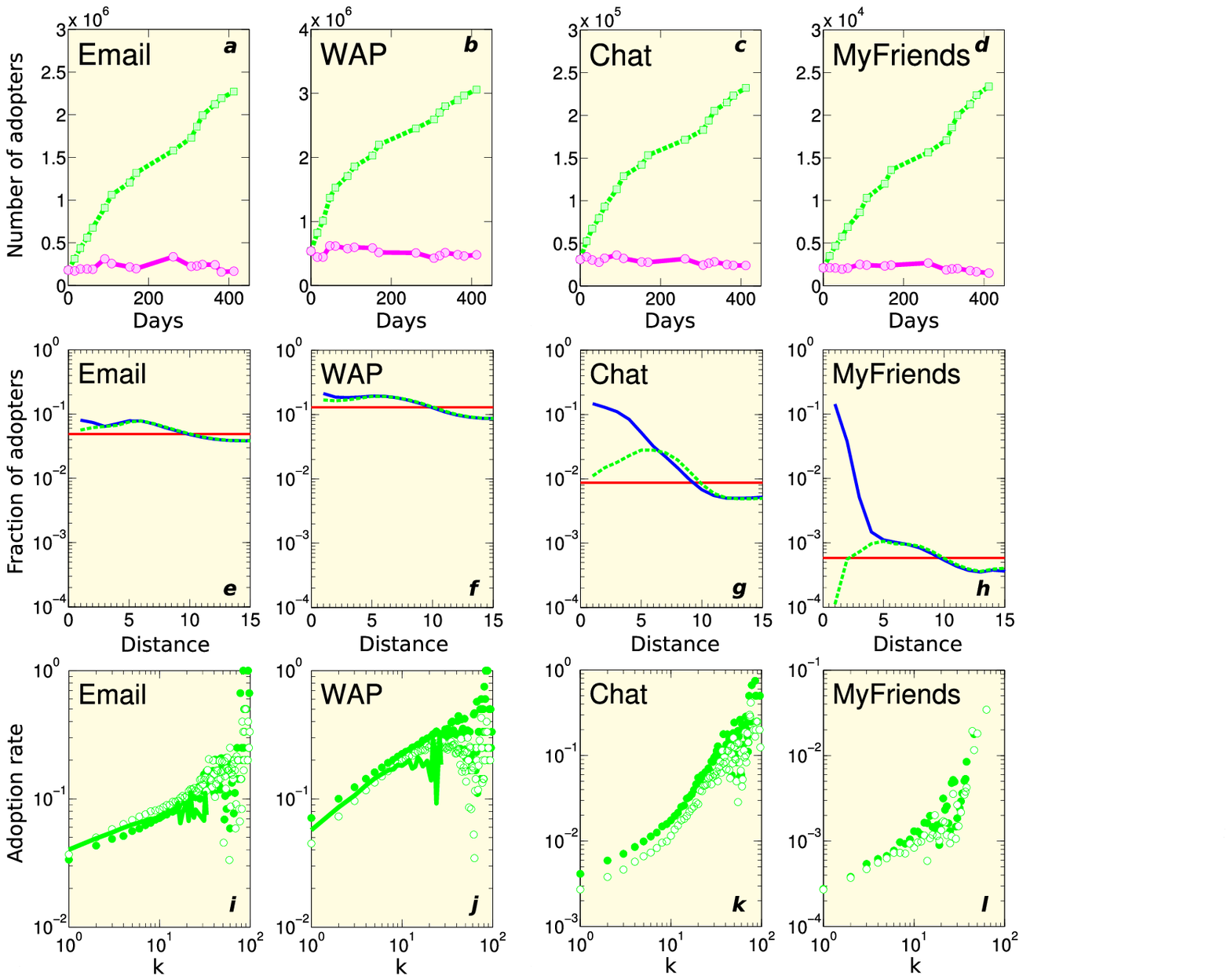}
\end{center}
\renewcommand{\baselinestretch}{1}\normalsize
\caption{Characterizing service adoption patterns. \textbf{(a--d)} The
number of users for each service in two week intervals (circles,
continuous), indicating that usage has reached a stationary value. The
cumulative number of users (squares, dashed) represent all those who
have used the service at least once since the beginning of our
observational period.  \textbf{(e--h)} The fraction of users at
distance $d$ from a service user (blue, continuous) or non-user
(green, dashed), averaged over 100 randomly chosen users
(non-users). The horizontal line indicates the expectation based on
random adoption ($p_i^2$). \textbf{(i-l)} Adoption rates in function
of the user degree $k$. Filled symbols stand for call data aggregated
over a one-month period, while empty ones denote a one-month period 10
months later. The solution of Eq.~(\ref{eq:stationary-sol}), valid for
Email and WAP, is shown as a solid line.
}
\label{fig-coverage-as-distance}
\end{figure*}

To check if service usage shows effects beyond the dyads, so that most
members of a community tend to adopt the same service, we determined
the fraction $\rho_i^\mathrm{U}(d)$ of users of service $S_i$ among
individuals $d$ links away from a user of $S_i$. For reference we also
measure the fraction $\rho_i^\mathrm{N}(d)$ of users in the vicinity
of an individual that has \emph{not} adopted $S_i$. If adopters do not
segregate into communities, then $\rho_i^\mathrm{U}(d) =
\rho_i^\mathrm{N}(d)$.  Indeed, we find that for Email and WAP
$\rho_i^{\mathrm{U}}(d)$ and $\rho_i^{\mathrm{N}}(d)$ largely overlap
for most $d$ values (Fig.~1 \emph{e} and \emph{f}), and the observed
difference between them at small $d$ is relatively small, suggesting
either the absence of compact user communities or at most weak local
aggregation of users.  In contrast, for Chat and MyFriends
$\rho_i^{\mathrm{U}}(d)$ is one to two orders of magnitude larger than
$\rho_i^\mathrm{N}(d)$ for small $d$
(Fig.~\ref{fig-coverage-as-distance} \emph{g} and \emph{h}),
indicating that the vicinity of a user has a significantly higher
density of users than the vicinity of a non-user. To illustrate the
presence or absence of a community-based segregation we visualized the
adoption patterns in selected neighborhoods for Email and Chat.  We
find that Email is used uniformly across communities, with no apparent
correlation between the underlying community structure and service
usage (Fig.~2\emph{a}). In contrast, Chat users have the tendency to
form compact communities, so that in a given community either almost
everybody or almost nobody uses Chat (Fig.~2\emph{b}).

If service usage is driven by communication with social contacts,
high-degree individuals should more likely use a given service than
small-degree individuals. While the role of hubs and connectors in
innovation diffusion is well
documented~\cite{rogers,bass,valente,coleman}, in the absence of
sufficient statistics the functional dependence of the adoption rate
on the degree remains unknown. Furthermore, it is unclear if such
degree-dependent effects should emerge in the usage patterns as
well. Therefore, we determined the usage rate among users with degree
$k$, finding that for each service the probability that an individual
uses a given service increases with her degree
(Fig.~1\emph{i}--\emph{l}). For example, the likelihood that an
individual with 100 contacts uses Chat is more than two orders of
magnitude higher than for an individual with one or two contacts.

In summary, the empirical results indicate that a common feature of
all four services is that they reached a highly dynamic saturated
state, characterized by a degree-dependent usage. Yet, there are also
significant differences between them, suggesting the existence of two
adoption classes: Chat and MyFriends display a strong community-based
segregation and high dyadic effects, in contrast with Email and WAP,
that are characterized by weaker dyadic effect and the apparent
absence of compact user communities. Thus for each service we observe
correlations between the structure of the underlying communication
network and the likelihood of service usage by individual users.

A methodological challenge in understanding service usage is that we
do not know why an individual uses a specific service: is she
persuaded by an acquaintance, or influenced by non-social factors,
such as service utility, an advertisement, or a newspaper article?
Furthermore, an individual can be persuaded to use a service by an
acquaintance that she never contacts through her mobile phone. To
account for these ambiguities, we grouped all individuals in two
categories: \emph{connected} (``\emph{c}'') individuals are those that
are in phone contact with at least one user of service $S_i$, thus
have a chance to be influenced by an acquaintance, and \emph{isolated}
(``\emph{i}'') individuals represent those that have no user of $S_i$
among their phone contacts, thus can be influenced only by external
factors, such as mass media or acquaintances they do not
call~\cite{lekvall}. Furthermore, we will call an individual a {\it
user} if during a one month interval she used service $S_i$, and a
\emph{potential user} if she did not use the service during the
month-long period, but could potentially become a user in the
future. Therefore, we distinguish between four categories of
individuals, isolated and connected potential users, as well as
isolated and connected users, the number of $k$-degree individuals at
time $t$ in each category being given by $N_k^i(t)$, $N_k^c(t)$,
$U_k^i(t)$ and $U_k^c(t)$, respectively. For a given service the
changes in the number of isolated and connected adopters with degree
$k$ can be written as
\begin{eqnarray}
\frac{\partial U_k^i(t)}{\partial t} & = & \mu_k N_k^i(t) - \eta_k
U_k^i(t)
\label{eq:delta_aik} \\
\frac{\partial U_k^{c}(t)}{\partial t} & = & \left\{ \lambda_k [1 +
  \Theta(t)(k - 1)] + \mu_k \right\} N_k^{c}(t) \\ \nonumber
& & - \xi_k U_k^{c} (t),
\label{eq:delta_anik}
\end{eqnarray}
where $\mu_k$ is the rate at which a potential user becomes an user
due to external influence, and $\eta_k$ ($\xi_k$) represents the rate
at which isolated (connected) users stop using the
service~\cite{parthasarathy}, going from users to potential users. For
connected potential users $\lambda_k$ gives the rate at which they
turn users as a result of the influence of one of their user contacts,
and $\Theta(t)$ is the probability that a user's randomly chosen link
points to a user, given
by~\cite{berthelemy,vespignani-virus,vespignani-internet}
\begin{eqnarray}
\Theta(t) & = & \sum_{k'} \frac{k' P(k')}{\langle k \rangle}
  \frac{U_{k'}(t)}{N P(k')},
\label{eq:theta}
\end{eqnarray}
where the first term in the product is the probability of finding a
degree-$k'$ individual at the end of a randomly chosen link if the
degree distribution of the communication network is $P(k)$, and the
second term is the probability that this individual is an user,
$U_{k'}(t) = U_{k'}^i(t) + U_{k'}^{c}(t)$, and $N$ is the number of
individuals in the network. The time evolution of the number of
potential users is given by
\begin{eqnarray}
\frac{\partial N_k^i(t)}{\partial t} & = & -\frac{\partial
  U_k^i(t)}{\partial t} - \Pi_k(t)
\label{eq:delta_uik} \\
\frac{\partial N_k^{c}(t)}{\partial t} & = & -\frac{\partial
  U_k^{c}(t)}{\partial t} + \Pi_k(t),
\label{eq:delta_unik}
\end{eqnarray}
where $\Pi_k(t)$ is the rate at which isolated potential users acquire
user neighbors, becoming connected. If degree correlations are absent,
we can assume that any non-user neighbor with degree $k'$ turns user
at the same rate $\frac{\partial U_{k'}(t)}{\partial t}
\frac{1}{N_{k'}(t)}$, where $N_{k'}(t) = N_{k'}^i(t) +
N_{k'}^{c}(t)$. Accordingly, for isolated potential users with degree
$k$
\begin{eqnarray}
\Pi_k(t) & = & N_k^i(t) \, k \sum_{k'} \frac{k' P(k')}{\langle k
  \rangle} \frac{\partial U_{k'}(t)}{\partial t} \frac{1}{N_{k'}(t)}.
\label{eq:pi}
\end{eqnarray}
Next we show that Eqs.~(\ref{eq:delta_aik})--(\ref{eq:delta_anik})
with the closures provided by (\ref{eq:theta})--(\ref{eq:pi})
accurately predict the dynamics of the observed service usage pattern.

The input of (\ref{eq:delta_aik}) and (\ref{eq:delta_anik}) are the
time-independent adoption functions $\mu_k$, $\eta_k$, $\lambda_k$,
and $\xi_k$, which quantify the response of the user base to the
service. These adoption functions were measured using a
self-consistent method described in the \emph{Supporting Text}, and
are shown in Fig.~2\emph{c}--\emph{j}. The measurements were performed
for two intervals ten months apart, finding in both cases the same
functional form (see the filled and empty symbols in
Fig.~2\emph{c}--\emph{j}), supporting our starting hypothesis that the
adoption functions are time-independent. To test the predictive power
of the proposed analytical model, we numerically integrated
Eqs.~(\ref{eq:delta_aik})--(\ref{eq:delta_anik}), starting with the
initial condition in which no users are present at time $t = 0$. The
integration predicts that the number of users rapidly increases
followed by saturation (Fig.~3\emph{a}--\emph{d}). Most important, in
the saturated state (\ref{eq:delta_aik}) and (\ref{eq:delta_anik})
reproduce not only the total number of users, but also the detailed
$k$-dependence of all user categories (Fig.~3\emph{e}--\emph{h}). The
inspection of the adoption functions and the analytical model suggests
the existence of two classes of services, that we discuss separately
below.

\emph{Individual Services}: The most striking common feature of Email
and WAP is the fact that for both services $\mu_k \gg \lambda_k$
(Fig.~2 \emph{c} and \emph{d}), \emph{i.e.}\ external influence, such
as mass media and advertisements, dominate over personal persuasion.
Furthermore, for these two services $\eta_k \approx \xi_k$, indicating
that there is little difference in the probability that an isolated or
a connected user suspends the service. For $\mu_k \gg \lambda_k$ and
$\eta_k = \xi_k$ Eqs.~(\ref{eq:delta_aik})--(\ref{eq:delta_anik})
reduce to
\begin{eqnarray}
\frac{\partial U_k(t)}{\partial t} & = & \mu_k \left[ N P(k) - U_k(t)
\right] - \eta_k U_k(t),
\label{eq:delta_ak_individual}
\end{eqnarray}
where $U_k(t)$ denotes the total number of users with degree $k$. In
the stationary state Eq.~(\ref{eq:delta_ak_individual}) predicts that
\begin{equation}
\frac{U_k(t \rightarrow \infty)}{N P(k)} = \frac{1}{1 +
  \frac{\eta_k}{\mu_k}},
\label{eq:stationary-sol}
\end{equation}
\emph{i.e.}\ the adoption rate is determined only by $\mu_k$ and
$\eta_k$. As shown in Fig.~1 \emph{i} and \emph{j}, the prediction
(\ref{eq:stationary-sol}) is in good agreement with the measured
$k$-dependent adoption rates for Email and WAP.

\emph{Cooperative Services}: In contrast with the individual services,
for Chat and MyFriends $\lambda_k \gg \mu_k$, so the influence of
social contacts is more important than external
influence. Furthermore, for these services $\eta_k \gg \xi_k$ (Fig.~2
\emph{i} and \emph{j}), implying that isolated users are more likely
to abandon the service than connected ones. Thus
Eq.~(\ref{eq:stationary-sol}) is not expected to hold, but the
integration of (\ref{eq:delta_aik})--(\ref{eq:delta_anik}) correctly
reproduces the $k$-dependent adoption rates observed in the real data
(Fig.~3 \emph{g} and \emph{h}). The emergence of communities observed
for Chat and MyFriends is also a consequence of the functional form of
the adoption functions. Indeed, with high $\lambda_k$, if an
individual has two or more user neighbors, his or her likelihood of
adoption increases, enhancing the pressure on the members of the same
community to adopt the same service. At the same time $\eta_k \gg
\xi_k$ leads to a preferential discontinuation by isolated
individuals, leading to the observed virtual absence of users in
communities in which users do not dominate, and leaving usage intact
only in communities where most individuals use the same service.

Therefore, the empirically observed two distinct usage patterns can be
reduced to the relative strength of $\lambda_k$ and $\mu_k$,
quantifying the user base's attitude to a specific service. Indeed,
Chat and MyFriends facilitate alternative ways of communication
between individuals, thus they are subject to network
externalities~\cite{farell,Katz}: the value of a service to a given
user increases if her acquaintances also adopt the service,
responsible for the observed $\lambda_k \gg \mu_k$ relation. Isolated
users do not experience such peer pressure, thus are more likely to
drop the service than users surrounded by users, responsible for the
fact that $\eta_k \gg \xi_k$. In contrast, we find that Email and WAP
have no immediate impact on the user's social contacts. This is
obvious for WAP, whose main purpose is to provide Web access, thus
does not require the cooperation of other users to be useful. The fact
that Email is an individual service is highly paradoxical, however,
given that Email is a communication service, just like Chat. There is
a major difference, though: Chat is an entirely phone-based service,
thus it is useless unless the phone contacts of a user also adopt
it. In contrast, for Email the mobile phone serves only as an
interface, which allows the user to communicate with all those who use
electronic mail, either using the phone or a computer (which is the
far most common case) to access it. Thus there is no need for all
phone-based acquaintances of a user to adopt Email in order to make it
useful for an individual phone user. Therefore, when it comes to
mobile phone usage, Email is not subject to network
externalities~\cite{farell,Katz} and increasing returns~\cite{arthur},
having the characteristics of an individual rather than a cooperative
service.

While the continuum theory proposed above provides an accurate
description of the observed usage patterns, it is of value only if it
offers predictions of practical importance. For many services and
products that are beneficial to a user base, a question of paramount
importance emerges: how can one enhance service usage? The standard
ways are advertisements and incentives, increasing awareness of the
service and boosting the user base. Such measures are costly and
therefore can be used only for a limited duration. They are fueled by
the belief that once users become familiar with a service, they will
likely continue its usage. Yet, next we show that in the saturated
regime such approaches are ineffective: once the incentive is removed,
the number of users returns to the pre-incentive value. Indeed,
Eq.~(\ref{eq:stationary-sol}) predicts that the number of users
depends only on $\eta_k$ and $\mu_k$ for individual
services. Incentives can temporarily increase the adoption rate
($\mu_k$) and decrease the dropout rate ($\eta_k$, $\xi_k$), but once
they are discontinued, the number of users ($U_k(t)$) returns to their
pre-incentive value. To see if this result applies to all services, we
used numerical simulations (see \emph{Supporting Text}) with the
adoption functions of Fig.~2\emph{c}--\emph{j} as input, finding that
if we artificially doubled the number of users, the usage returns to
its original value for both individual and cooperative services (see
Fig.~4\emph{a}). The final evidence comes from two services, providing
collect calls and text message storage, available to the same user
base as the four services studied above.  These services were promoted
by campaigns that started after they have reached saturation. As
Fig.~\ref{fig:predictions} \emph{c} and \emph{d} show, the campaigns
have resulted in more than a five-fold increase in the number of
users, but both services returned to their saturation value once the
campaigns ended. Taken together, we find that once the saturated
regime has been reached, traditional methods for boosting the consumer
base are only effective for a limited time. Yet, next we show that for
cooperative services one can exploit the intimate knowledge of the
communication network to design cost-effective incentives.

Highly connected individuals or hubs exert disproportional influence
over the user base~\cite{valente2}. Therefore, if each user with
degree exceeding $k_c$ is given permanent free usage of service $S_i$,
increasing their likelihood of adoption ($\mu_k$) and a decreasing
their dropout rate ($\eta_k$ and $\xi_k$), such differentiated
incentives will result in higher saturation values (see Fig.~4
\emph{e} and \emph{f}). Indeed, those who received the incentive will
increase usage, an effect that applies to both individual and
cooperative services. However, due to the significant $\lambda$ term
for cooperative services, increased usage by hubs will also enhance
the usage of their acquaintances, an effect which will be negligible
for individual services. The question is, will this second, persuasion
induced effect make the incentive potentially cost effective,
recovering the cost of providing free permanent usage for all hubs? To
answer this we determined the impact of the incentive by calculating
the ratio between the number of new users (the difference $\Delta$
between the pre- and post-incentive saturation values) and the number
of hubs $N_h(k \ge k_c)$ that have been provided the incentive. As
Fig.~4\emph{b} shows, we find that the impact increases with $k_c$,
indicating that the larger the degree of the hubs the incentives were
given to, the larger their potential impact. For individual services
($\lambda = 0$), however, the impact is negligible and has only weak
dependence on $k_c$ (Fig.~\ref{fig:predictions}\emph{b}). If one
factors in the potential cost of the incentives, we find that
hub-oriented permanent incentives could benefit cooperative services,
given that for these the impact increases with $k_c$, thus for any
incentive cost there will be a $k_c$ value for which the lost revenue
on the hubs can be recovered by the increased revenue from the users
that have adopted under the influence of the hubs. However, for
individual services a hub-based incentive would likely be a losing
proposition: the revenue lost from the heavy usage of the hubs cannot
be recovered from the increased usage of their contacts, as for these
services the impact of persuasion is negligible.

In summary, while informational externalities, capturing the spread of
information about a given service between acquaintances, are expected
to play an important role in the adoption of innovations, their role
is diminished in the saturated regime, where awareness of the benefits
of a given service are widely known. We would expect, therefore, that
social network effects should cease to correlate with service
usage. Here we provide direct evidence of the contrary: for all
services social network affects can be detected, even in the
stationary state. The magnitude of this effect is service-dependent,
however, leading to the coexistence on the same social network of two
distinct classes of services. We find that services characterized by
strong dyadic and community effects are driven by network
externalities, making the use of a service more valuable for a user if
it is used by its acquaintances as well. Most importantly, we find
that successful interventions into service usage require a detailed
understanding of the interplay between the structure of the underlying
communication network and the usage mechanism. There is no one-size
fits all advertising strategy: for a given innovation some strategies
are guaranteed to fail, while strategies based on the quantitative
knowledge of the spreading mechanism are more likely to
succeed. Finally, our approach offers a coherent framework for
understanding service usage in an environment where both the social
network and service usage can be monitored, potentially opening new
avenues for research into social diffusion and human dynamics.

\begin{figure*}
\begin{center}
\includegraphics[width=0.8\textwidth]{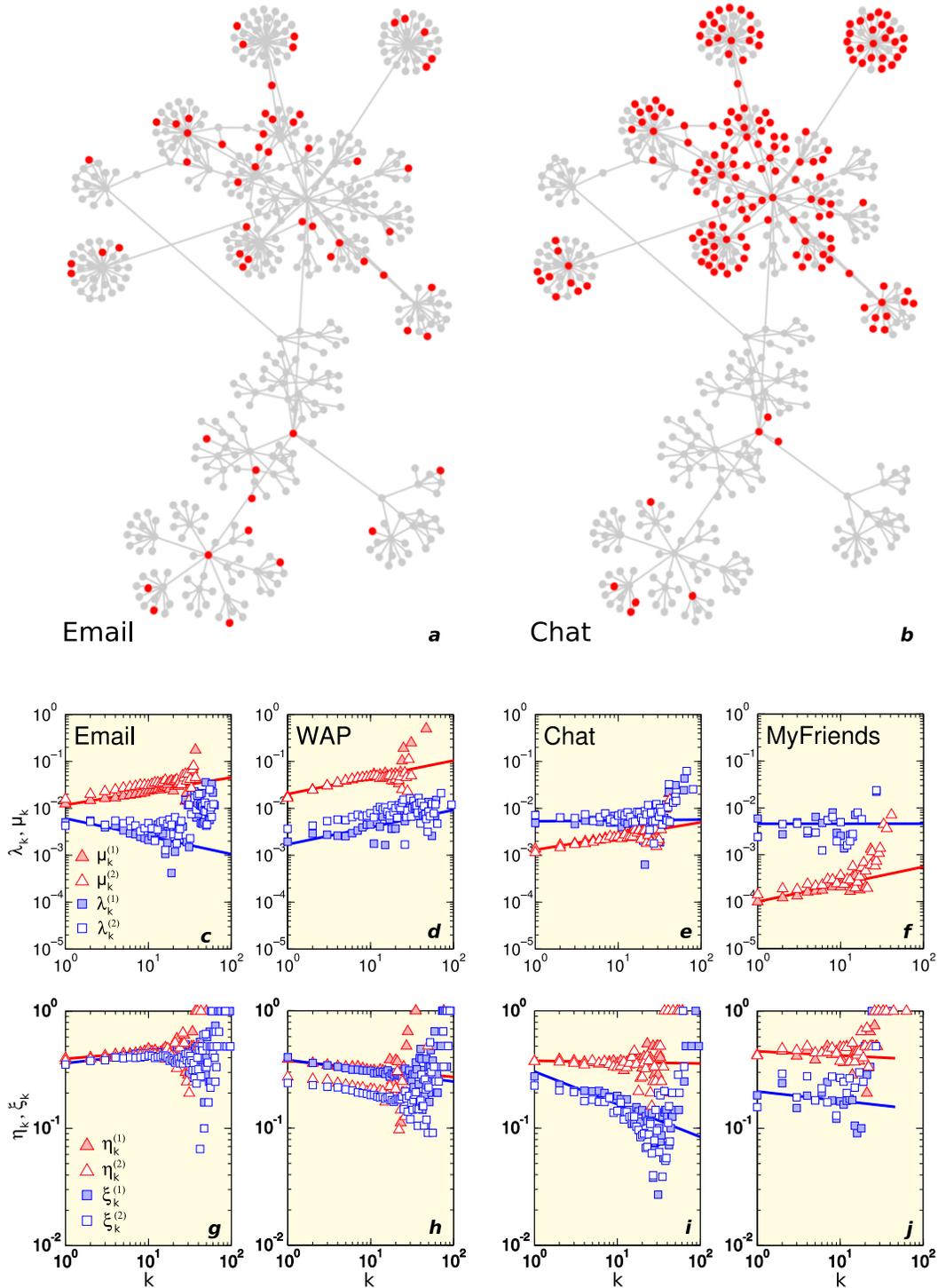}
\end{center}
\renewcommand{\baselinestretch}{1}\normalsize
\caption{A local neighborhood of the communication network of
telephone contacts. \textbf{(a), (b)} In both figures each node
corresponds to an individual (mobile phone number), two individuals
being connected if they talked or sent short messages to each other
during a one-month period. The network was drawn by selecting an
individual, and displaying all nodes within distance four from her. To
illustrate service usage on this network, Email users are colored red
in (a), while in (b) Chat users are shown in red. Note that while
Email users are scattered uniformly, Chat users form compact
communities. \textbf{(c--f)} The degree-dependent adoption functions
characterizing the adoption of services thanks to external influence
($\mu_k$) and the influence of service users with direct contact to
the user ($\lambda_k$) (see \emph{Supporting Text}). For each adoption
function we show the results of two independent measurements (filled
and empty symbols), made 10 months apart (the two intervals denoted by
the upper indices in the legends), indicating that the functional form
of the adoption functions is time-independent. The vertical shifts
represent fluctuations in the coverage (see Fig.~1\emph{a}--\emph{d}),
and do not affect our results. \textbf{(g--j)} Adoption functions
capturing the service suspension by isolated ($\eta_k$) or connected
($\xi_k$) users, again measured for two different time intervals.}
\label{fig:ratescomparison}
\end{figure*}

\begin{figure*}
\begin{minipage}[h]{\linewidth}
\begin{center}
\includegraphics[width=0.8\textwidth]{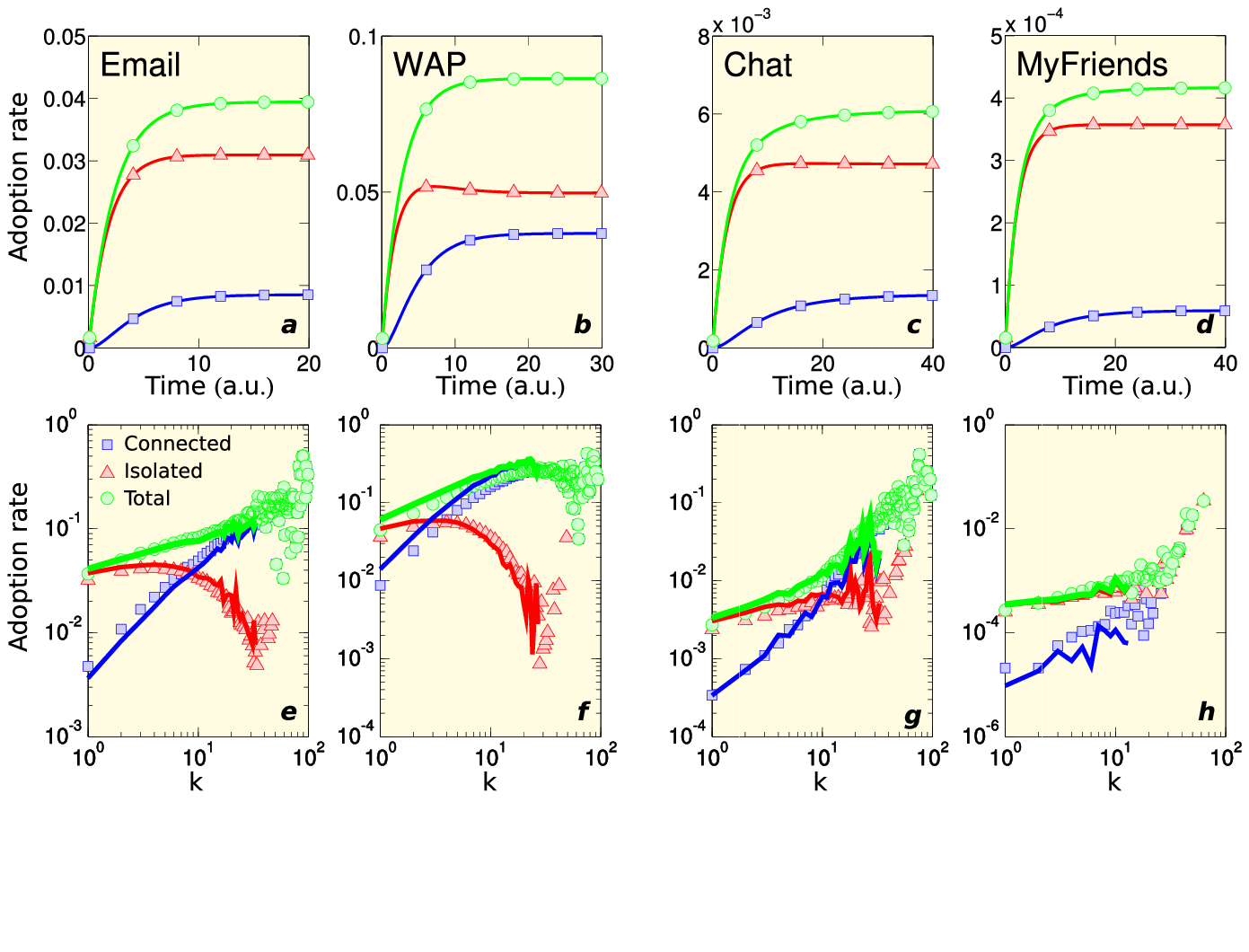}
\end{center}
\end{minipage}
\renewcommand{\baselinestretch}{1}\normalsize
\caption{Predicted and measured adoption rates. \textbf{(a-d)} The
growth of the adoption rates (fraction of all users that use the
specific service) as predicted by the integration of
Eqs.~(1)--(2). The adoption rate of isolated (connected) users is
shown as triangles (squares), and the sum of the two, corresponding to
the total rate of adoption, as circles. \textbf{(e-h)} The predicted
adoption rates for one month using
Eqs.~(\ref{eq:delta_aik})--(\ref{eq:pi}) and the adoption functions in
Fig.~\ref{fig:ratescomparison}\emph{c}--\emph{j}, with solid lines,
and the actual measurements with symbols. The fraction of isolated
users is shown as triangles, that of connected ones with squares, and
the sum of the two with circles. \label{fig:parameters} }
\end{figure*}

\begin{figure*}
\begin{center}
\includegraphics[width=0.5\textwidth]{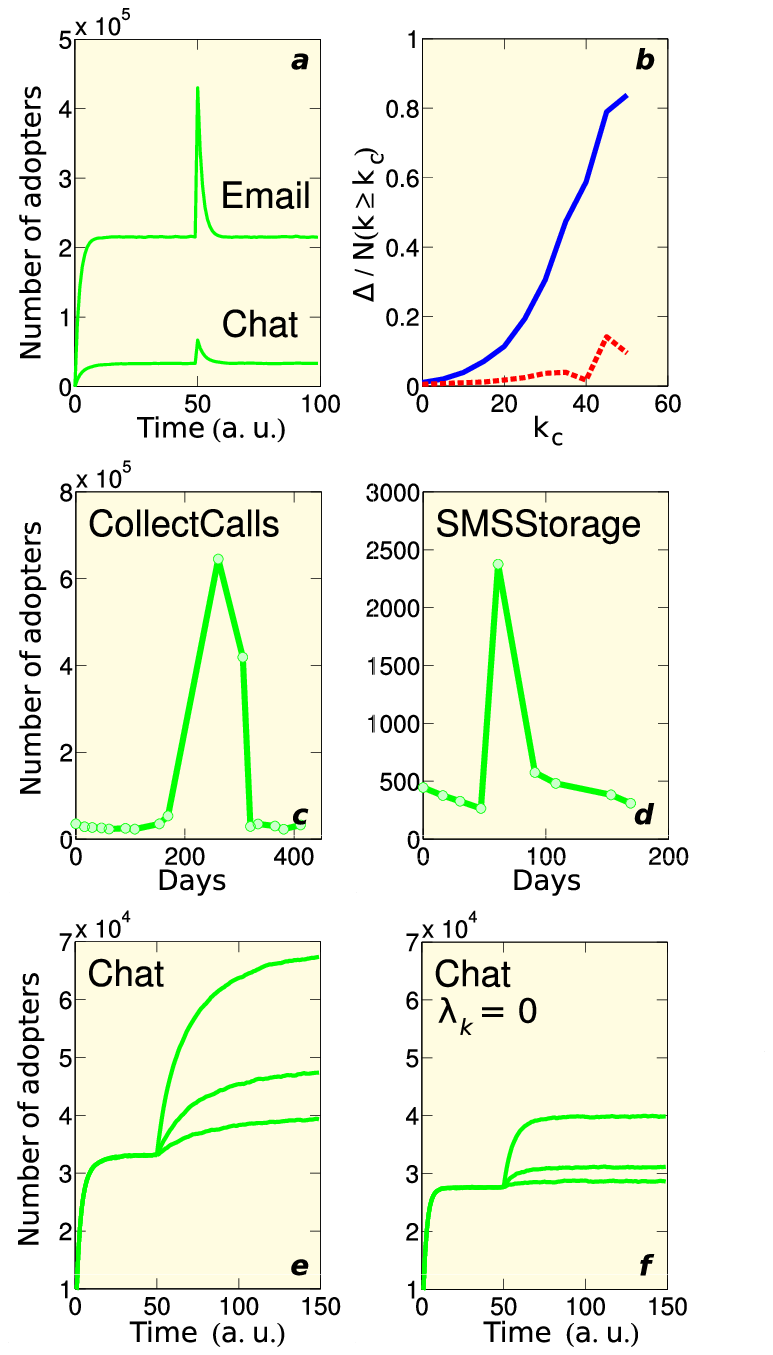}
\end{center}
\renewcommand{\baselinestretch}{1}\normalsize
\caption{The impact of advertising and targeted interventions on the
number of service users. \textbf{(a)} The changes in the number of
users after it was doubled by a short-term effect, such as
advertising, for Email and Chat.  \textbf{(b)} The gain in the number
of users $\Delta$ in the stationary states (see \emph{e} and
\emph{f}), normalized by the number of subscribers who have been
offered permanent incentives ($N(k \ge k_c)$). The continuous curve
corresponds to Chat, while the dashed one has the same parameters as
Chat, but with $\lambda_k = 0$. \textbf{(c), (d)} The number of
service users for two services, for which the provider has boosted the
number of users using advertising and incentives. \textbf{(e)} The
changes in the number of adopters for Chat after hub-based incentives were
initiated at $t=50$, by decreasing in half $\eta_k$ and $\xi_k$ for
hubs with degree over 5, 10, and 15, for the three curves from the top
to bottom, respectively. \textbf{(f)} Number of adopters as in \emph{e},
with the adoption function $\lambda_k$ set to 0 at the beginning of
the simulation, to mimic the impact of the lack of personal
persuasion.
\label{fig:predictions} }
\end{figure*}

\clearpage

\onecolumngrid
\begin{center}
\textbf{\large Supporting Text for ``Network effects in service usage''}
\vspace{0.5cm} \\
G\'abor Szab\'o and Albert-L\'aszl\'o Barab\'asi
\end{center}
\vspace{1.5cm}
\twocolumngrid

\section{Datasets}

Based on the call records of a mobile telephone operator with
residential subscribers, we performed our study on data comprising
communication spanning 14 months, starting April 15, 2004. We
neglected calls whose destination or origin was outside the company,
which left us with approximately 5.5 million nodes in the call network
for each month. The links between nodes were defined by initiated
voice calls and short messages (SMS), the two most common services
with origin and destination information. While these are directed by
nature, we made the communication graph undirected, yielding more than
10 million links for one month. The call data are aggregated into
intervals of two weeks, each record showing the caller, the callee,
the type of the communication, and its frequency during the two
weeks. For call graphs, we accumulated two of the datasets together
such that two users are linked if they communicated at least once
during the two weeks (see Table~\ref{table:periods} for the time
intervals). While the overwhelming majority of individuals only use
voice calls for communications, it was possible to identify users of
the studied services based on additional fields present in the
datasets or by checking the callee field for unique service numbers
that need to be dialed to use the service.

\section{Measuring the adoption functions $\mu_k$, $\eta_k$,
  $\lambda_k$, and $\xi_k$}

\begin{figure}
\begin{center}
\includegraphics[width=0.6\linewidth]{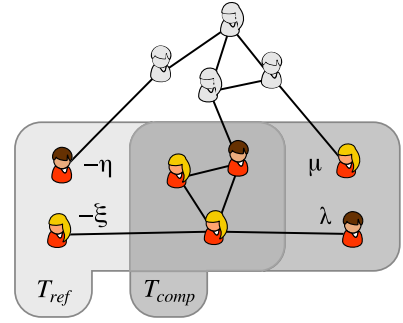}
\end{center}
\renewcommand{\baselinestretch}{1}\normalsize
\caption{Schematic representation of the mobile phone users' network.
Users of a service are shown on the bottom. The estimation of the
adoption functions $\mu_k^*$, $\eta_k^*$, $\lambda_k^*$, and $\xi_k^*$
is based on two time windows of 4 weeks each with a 2-week overlap,
represented by the two gray boxes ($T_{\mathit{ref}}$ and
$T_{\mathit{comp}}$, respectively). Users only in $T_{\mathit{ref}}$
but not in $T_{\mathit{comp}}$ allow us to determine the terms in
Eqs.~(\ref{eq:discr_delta_aik})--(\ref{eq:discr_delta_anik}),
responsible for the temporary suspension of the service, \emph{i.e.}
$-\eta$ for isolated discontinuation and $-\xi$ for connected
discontinuation. Individuals who are users in $T_{\mathit{comp}}$ only
(they were not users before $T_{\mathit{comp}}$) are responsible for
the gain in the rate equations, \emph{i.e.} the $\mu$ and $\lambda$
terms for isolated and connected new users, respectively.
\label{fig:time-window}
}
\end{figure}

While Eqs.~(1) and (2) are continuous in time, in order to be able to
measure the adoption functions, we need to consider their discretized
versions, given that the datasets for service usage are aggregated for
periods of two weeks. Therefore, we write the discretized adoption
equations as
\begin{eqnarray}
\Delta u_k^i(T_{\mathit{comp}}, T_{\mathit{ref}}) & = & u_k^{i,+} -
u_k^{i,-} \\ \nonumber
& = & \mu_k^* n_k^i(T_{\mathit{ref}}) - \eta_k^*
u_k^i(T_{\mathit{ref}})
\label{eq:discr_delta_aik} \\
\Delta u_k^c(T_{\mathit{comp}}, T_{\mathit{ref}}) & = & u_k^{c,+} -
  u_k^{c,-} \\ \nonumber
& = & \left\{ \lambda_k^* u_k^{nn} + \mu_k^* \right\}
  n_k^c(T_{\mathit{ref}}) - \xi_k^* u_k^c(T_{\mathit{ref}}),
\label{eq:discr_delta_anik}
\end{eqnarray}
where we denoted with $u_k^i$, $u_k^c$, $n_k^i$, and $n_k^c$ the
measured values of the variables $U_k^i$, $U_k^c$, $N_k^i$, and
$N_k^c$, respectively. $T_{\mathit{ref}}$ and $T_{\mathit{comp}}$
denote two time frames with data for both service usage and the
communication map between the subscribers. The network of social
contacts was measured with respect to $T_{\mathit{ref}}$: two
individuals were considered to be acquaintances if they called or sent
a short message to each other in the $T_{\mathit{ref}}$ time
period. An isolated (``$i$'') individual was an individual without a
service user in the immediate neighborhood in $T_{\mathit{ref}}$,
while a connected (``$c$'') individual had at least one user among its
acquaintances. A schematic representation of the measurement procedure
is shown in Fig.~\ref{fig:time-window}.

We first focus on Eq.~(\ref{eq:discr_delta_aik}), which describes the
change in the number of isolated users with degree $k$ during the
$T_{\mathit{comp}}$ time interval, $\Delta u_k^i(T_{\mathit{comp}},
T_{\mathit{ref}})$, where the reference social network and usage data
was taken from $T_{\mathit{ref}}$. The change is the result of a gain
in users due to isolated adoption ($u_k^{i,+}$), and a loss of these
through isolated service abandoning ($u_k^{i,-}$), both of which can
be experimentally measured: $u_k^{i,+}$ is the number of new isolated
users in $T_{\mathit{comp}}$ with degree $k$ who were not users during
$T_{\mathit{ref}}$, and $u_k^{i,-}$ gives the number of isolated users
in $T_{\mathit{ref}}$ who in turn abandoned the service in
$T_{\mathit{comp}}$. Since these processes are precisely what
$\mu_k^*$ and $\eta_k^*$ aim to capture, we can determine the two
adoption functions using
\begin{eqnarray}
\mu_k^* & = & \frac{u_k^{i,+}}{n_k^i(T_{\mathit{ref}})}
\\
\eta_k^* & = & \frac{u_k^{i,-}}{u_k^i(T_{\mathit{ref}})},
\end{eqnarray}
where $u_k^i(T_{\mathit{ref}})$ and $n_k^i(T_{\mathit{ref}})$ are the
number of isolated users and non-users with degree $k$ in
$T_{\mathit{ref}}$, respectively.

Connected potential users in $T_{\mathit{ref}}$
($n_k^c(T_{\mathit{ref}})$) are also influenced to adopt by
interpersonal communication, apart from external channels. Also,
connected users ($u_k^c(T_{\mathit{ref}})$) may also abandon the
service, allowing us to measure the two other adoption functions
$\lambda_k^*$ and $\xi_k^*$:
\begin{eqnarray}
\lambda_k^* & = & \left[ \frac{u_k^{c,-}}{n_k^c(T_{\mathit{ref}})} -
\mu_k^* \right] / u_k^{nn} \\
\xi_k^* & = & \frac{u_k^{c,-}}{u_k^c(T_{\mathit{ref}})},
\end{eqnarray}
where $u_k^{nn}$ is the measured \emph{average exposed degree} for
connected potential users, \emph{i.e.}\ the average number of user
neighbors for degree-$k$ connected individuals (note that $u_k^{nn}
\ge 1$), and it is approximated by $1 + \Theta(t)(k - 1)$ in the rate
equation (2).

\begin{table*}
\begin{tabular}{|c||c|c|c|}
\hline
Adoption functions in Fig.~2\emph{c}--\emph{j} &
$T_{\mathit{ref}}$ & $T_{\mathit{comp}}$ & social network \\
\hline \hline
filled symbols & 4/15/2004--5/15/2004 & 5/1/2004--6/1/2004 &
4/15/2004--5/15/2004 \\
empty symbols   & 2/15/2005--3/15/2005 & 3/1/2005--4/1/2005 &
2/15/2005--3/15/2005 \\
\hline
\end{tabular}
\renewcommand{\baselinestretch}{1}\normalsize
\caption{The observation periods chosen for the measurement of the
adoption functions, as shown in Fig.~2\emph{c}--\emph{j}. The fourth
column shows the period in which the network of social contacts was
measured.
\label{table:periods}
}
\end{table*}

The observation periods used in this article are shown in
Table~\ref{table:periods}; both $T_{\mathit{ref}}$ and
$T_{\mathit{comp}}$ span one month, and they overlap by two weeks. On
one hand, the overlap was necessary, because we wished to make
$T_{\mathit{ref}}$ and $T_{\mathit{comp}}$ long enough so that
considerable adoption can be observed, and the longest consecutive
time frame for which we have communication data is one and a half
months. On the other hand, it demonstrates that the choice of
$T_{\mathit{ref}}$ and $T_{\mathit{comp}}$ is arbitrary, as long as
the adoption functions are understood to refer to the chosen time
window span and the relative shift between $T_{\mathit{comp}}$ and
$T_{\mathit{ref}}$. Therefore, the adoption functions shown in
Fig.~2\emph{c}--\emph{j} describe adoption and abandoning
probabilities for individuals observed during one month, with respect
to the next month two weeks into the future. The adoption rates also
have to be measured accordingly, as has been done in
Fig.~2\emph{c}-\emph{j} for $T_{\mathit{ref}}$ of the 2005 data (see
also Table~\ref{table:periods}).

The interpretation of the adoption functions is the following: an
individual who at least once uses the service during the one-month
observation period is called an \emph{user}. If we were able to
monitor service usage continuously in time, the number of users at any
point in time would follow a continuous curve. The number of users at
time $t$, in this case, would give the number of individuals who used
the service in the one-month period surrounding $t$. Since service
usage is provided to us in two-week intervals, we can only sample this
function at discrete intervals, such as $T_{\mathit{ref}}$ and
$T_{\mathit{comp}}$. While at any point in time the adoption functions
$\mu_k$, $\eta_k$, $\lambda_k$, and $\xi_k$ would be calculated as the
slope of their respective functions giving the number of new users or
abandonings, we are in effect approximating them by the slope of the
line connecting the discrete sampled values.

\section{Time independence of the adoption functions}

A fundamental assumption of our modeling framework is that the
adoption functions $\mu_k$, $\lambda_k$, $\eta_k$ and $\xi_k$ are
time-independent, and they characterize the user base's attitude to a
specific innovation. Thus it is crucial to test if this time
independence is indeed true in the available dataset. For this we have
measured the adoption rates and the adoption functions independently
for two different time intervals that are ten months apart from each
other (see Table~\ref{table:periods}). The results of the measurements
obtained for the first interval are shown as filled symbols in
Fig.~1\emph{i}--\emph{l} and Fig.~2\emph{c}--\emph{j}, respectively,
while those from the second interval are shown as empty symbols. As
one can notice, the functional forms of the adoption curves remain
unchanged in time. The only noticeable systematic difference is the
vertical shifts in the curves, while leaving the functional form
unchanged. These vertical shifts are a simple consequence of the
fluctuations in the service usage observed in
Figs.~1\emph{a}--\emph{d}. These measurements thus validate our
assumption that the only significant time dependence comes in the
number of users, as described by Eqs.~(1)--(2), and at those time
scales the adoption functions can be taken to be time independent.

\section{Integrating the rate equations}

\begin{figure}
\begin{center}
\includegraphics[width=0.8\linewidth]{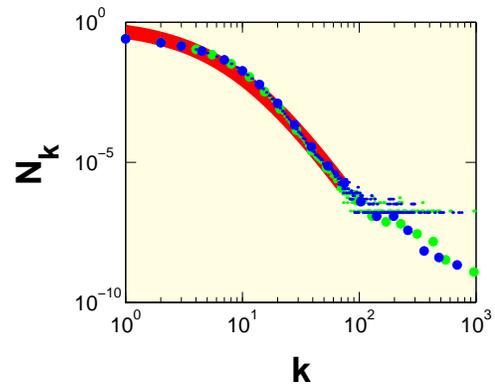}
\end{center}
\renewcommand{\baselinestretch}{1}\normalsize
\caption{The degree distributions of the two call networks in the
reference time windows of Fig.~2\emph{c}--\emph{j}, respectively. Data
from April 2004 are shown in green, and those from February 2005 in
blue.  Logarithmic binning was applied to the raw data, as shown by
the larger dots. A function of the form $A (k + k_0) ^ {-\gamma}$ was
fitted to the data points (thick red line), with parameters $k_0 =
9.3$ and $\gamma = 6.1$. The distributions overlap to a high degree,
indicating the time-independent nature of $P(k)$.}
\label{fig:degree-distributions}
\end{figure}

Once the adoption functions have been measured, it was possible to
predict the time evolution of the number of users by numerically
integrating the rate equations (1)--(8), using the measured adoption
functions $\mu_k$, $\eta_k$, $\lambda_k$, and $\xi_k$. As input we
took the degree distribution of the social network at
$T_{\mathit{ref}}$, which is rather stable in time as
Fig.~\ref{fig:degree-distributions} indicates. We set an initial
condition in which we had no service users in the system at $t = 0$,
\emph{i.e.} $U_k^i(t = 0) = U_k^c(t = 0) = N_k^c(t = 0) = 0$, and
$N_k^i(t = 0) = N_k$, where $N_k$ denotes the number of individuals
with degree $k$.  We chose a sufficiently small value for the time
step ($\Delta t = 0.1$), and by Euler integration we determined the
adoption rates for the desired range in time.

\section{Numerical simulations}

The continuum theory (1)--(2) represents an approximation to the real
spreading process: it cannot account for all the details of the
community formation process, so when services start to percolate
(higher coverages) it will likely result in deviations from the real
process. Therefore, to test the validity of the model, as well as the
impact of advertisements (which in the short term result in high
service adoption levels) we have carried out numerical simulations of
the usage of the services on the social network of $T_{\mathit{ref}}$
in 2004. For this, we fitted power laws of the form $A k^B$ on the
adoption functions of Fig.~2\emph{c}--\emph{j} using a least-squares
fit, and used them as the empirical adoption functions in the
simulations. For example, for Chat, the following fits were used (see
Fig.~2\emph{c}--\emph{j}, straight lines):
\begin{eqnarray*}
\mu_k & = & 0.0013 k ^ {0.2942} \\
\eta_k & = & 0.3751 k ^ {-0.0115} \\
\lambda_k & = & 0.0053 k ^ {0.0174} \\
\xi_k & = & 0.3038 k ^ {-0.2789}.
\end{eqnarray*}
We then initialized every individual in the network to be a non-user
of the service, and we implemented a Monte-Carlo simulation using as
input a close approximation to the empirically measured adoption
functions. In every time step we updated each individual's service
usage state by comparing the appropriate adoption function value to a
randomly drawn number from a uniform distribution between 0 and
1. Thus for isolated (connected) users $\eta_k$ ($\xi_k$), and for
isolated (connected) potential users $\mu_k$ ($\lambda_k$) was used
according to their degree. We used parallel update to keep track of
the changes in the states of individuals.

To simulate the effect of campaigns and advertisements, upon reaching
saturation at $t = 50$, we determined the number of users in the
system. We then changed the states of the same number of randomly
chosen non-users so that they became users, thus simulating that the
service usage was artifically increased by a factor of two as a result
of the campaigns. After this, we let the system relax and reach
equilibrium again (Fig.~4\emph{a}).

To study the effect of targeted campaigns, at $t = 50$ we halved the
adoption functions $\eta_k$ and $\xi_k$ for every person with a degree
greater or equal to the preset critical degree $k_c$, thus simulating
a permanent change in the behavior for these individuals: they would
be less likely to abandon services than before. We then allowed the
system to reach the stationary state again.

In addition, where mentioned, we repeated the simulations several
times and averaged the time-dependent adoption rates or other
quantities to minimize the effect of fluctuations.

\begin{figure}[h!]
\begin{center}
\includegraphics[width=0.8\linewidth]{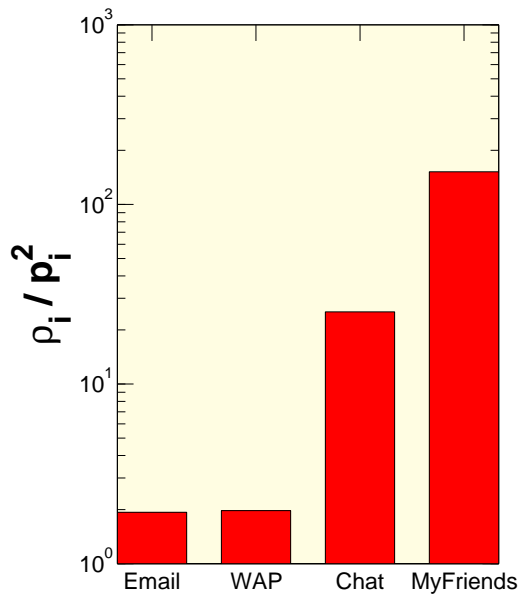}
\end{center}
\renewcommand{\baselinestretch}{1}\normalsize
\caption{The probability $\rho_i$ that a link connects two service
users, divided by the random expectation $p_i ^ 2$ when every
individual in the system adopts with the same probability $p_i$. $p_i$
is the overall measured penetration of the service.}
\label{fig:p11_histo}
\end{figure}

\end{document}